\documentclass[aps,prl,twocolumn,showpacs,groupedaddress]{revtex4}

\usepackage{graphicx}
\usepackage{dcolumn}
\usepackage{bm}

\begin{document}


\title{Nucleation and Collapse of the Superconducting Phase in Type-I Superconducting Films.}

\author{C. Gourdon}
\email{catherine.gourdon@insp.jussieu.fr}
\homepage{http://www.insp.upmc.fr}
\author{V. Jeudy}
\affiliation{
Institut des Nanosciences de Paris, CNRS~UMR~7588, Universit$\acute{e}$s~Paris~6~et~7\\ 
Campus Boucicaut, 140 rue de Lourmel, 75015 Paris - France}
\author{A. C\={e}bers}
\affiliation{
Institute of Physics, University of Latvia, Salaspils-1, LV-2169 - Latvia}
\date{\today}

\begin{abstract}
The phase transition between the intermediate and normal states in type-I superconducting films is investigated using magneto-optical imaging. Magnetic hysteresis with different transition fields for collapse and nucleation of superconducting domains is found. This is accompanied by topological hysteresis characterized by the collapse of circular domains and the appearance of lamellar domains. Magnetic hysteresis is shown to arise from supercooled and superheated states. Domain-shape instability resulting from long-range magnetic interaction accounts well for topological hysteresis. Connection with similar effects in systems with long-range magnetic interactions is emphasized.
\end{abstract}

\pacs{74.25.Ha, 05.65.+b, 75.70.Kw}

\maketitle

When submitted to a magnetic field, a type-I
superconductor undergoes a first-order phase transition between the superconducting (SC) and the normal-state (NS) homogeneous phases (HP). In the case of films in a perpendicular field, the transition proceeds through the onset of a modulated phase (MP), the so-called intermediate state (IS), which consists of an intricate pattern of SC and NS domains \cite{HKR,FHK,faber58,cebersPRB}. Such a transition is encountered in a variety of quasi-two dimensional systems
including ferromagnetic thin films with strong uniaxial
anisotropy \cite{hubertschafer,molho}, Langmuir polarized monomolecular layers at
air-water interface \cite{seul-chen}, magnetic fluids in Hele-Shaw
cells \cite{bacri}. MPs arise from the competition between short-range interactions associated with positive interface energy and long-range magnetic or dielectric interactions. Although these interactions have been recognized as a major ingredient in the description of the dynamics of pattern formation \cite{langer,dorsey,cebers2,cebersPRB}, they were not taken into account in former studies of interface motion \cite{frahm,liu}. An important issue concerning MP systems is the role of long-range interactions in the nucleation process of one phase into the other.
In a closely related field, the ion-induced nucleation of the liquid phase in the gas phase of a polar fluid, a problem which dates back to the invention
of Wilson's cloud chamber, this question is still under active debate \cite{onuki}.

In type-I SC films, it is well known that the SC-NS transition occurs at a magnetic field smaller than the bulk thermodynamical critical field $H_c$. This is due to the SC-NS interface energy and the magnetic stray field energy of the NS domains. The transition field was estimated using an approximate expression of this magnetic energy \cite{tinkham}. A more accurate prediction of the transition field should be obtained in the framework of the recently developed current-loop (CL) models \cite{dorsey,jeudy,cebersPRB} that made possible the calculation of the magnetic energy for various domain patterns. However, in SC films \cite{HKR,egorov}, as well as in magnetic systems \cite{molho}, a hysteresis loop opens up very close to the MP-HP boundary. Two distinct transition fields are found for the appearance and collapse of domains. Surprinsingly, the origin of this magnetic hysteresis still remains an open question. Does it arise from the existence of supercooled (SCL) and superheated (SH) metastable states? What is precisely the role of pinning centers and defects? Metastable states were clearly identified in dispersions of micron size SC spheres where the small volume reduces the probability of heterogeneous nucleation at defects \cite{feder,smith}. On the opposite, in large size systems like films, SCL and SH states are not expected to be observed \cite{FHK}. In addition to magnetic hysteresis, domain-shape hysteresis is found: domain shape and pattern are different for rising and decreasing field. The interplay between this topological hysteresis and the magnetic hysteresis at the MP-HP boundary is not well understood. Valuable insights into this question are expected from the study of the stability range of different domain shapes, a task that recent models have made possible \cite{cebersPRB,dorsey}.

In this Letter, we discuss the physical origin of magnetic and topological hysteresis close to the MP-HP boundary. The two transition fields are shown to correspond to the rupture of metastable states. They are used to determine the
critical radius for the nucleation and collapse of the SC phase. We show that topological hysteresis very likely originates from domain-shape instability arising from long-range interactions. The theoretical analysis of metastable states and domain-shape instabilities is carried out in the framework of the constrained current-loop (CCL) model which was successfully developed to take into account screening by superconducting currents in SC films \cite{cebersPRB}. 

The IS domain pattern in SC films was studied with the high-resolution Faraday microscopy imaging technique \cite{gourdon}. 
The samples consisted of indium films with thicknesses 0.6, 1.1, 1.5, 2.2, 10.0 $\pm\;$0.1 $\mu$m and 33 $\pm\;$3 $\mu$m. They were placed in an immersion-type cryostat in pumped liquid helium. The optical setup is similar to a reflection polarizing microscope \cite{cebersPRB}. The samples were zero-field cooled then subjected to a perpendicular magnetic field. 

\begin{figure}[tbp]
		\includegraphics [width=0.38\textwidth]{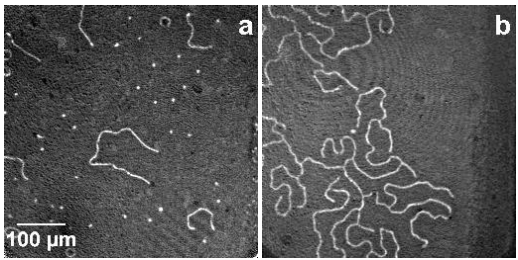}
	\caption{Intermediate state pattern in the 10 $\mu$m thick film close to the transition to the normal state. The superconducting (SC) domains appear in white and the normal phase in gray. (a) In rising magnetic field, SC bubbles and lamellae are observed ($H/H_c$=0.77 at $T$=1.92 K with $H_c(T=0)$=282 G), (b) In decreasing field, only lamellar domains are observed ($H/H_c$=0.68, $T$=1.85 K).}
	\label{fig:imagesIS}
\end{figure}

The SC-NS phase transition is found to be hysteretic. Distinct transition fields are found in rising and decreasing applied field. They are associated with different morphologies of the SC domains. Figure \ref{fig:imagesIS} shows typical IS patterns in the 10 $\mu$m thick film near the MP-HP boundary. In rising field lamellar and bubble-shape SC domains are observed (Fig. \ref{fig:imagesIS}(a)). The length of SC lamellae decreases until they are reduced to SC bubbles. The transition to the homogeneous NS phase results from the collapse of these bubbles whose diameter is 6-7 $\mu$m. In contrast, with decreasing field, the lamellar pattern appears abruptly in a very narrow range of field (Fig. \ref{fig:imagesIS}(b)). 
\begin{figure}[htbp]
	\begin{center}
		\scalebox{0.40}{\includegraphics{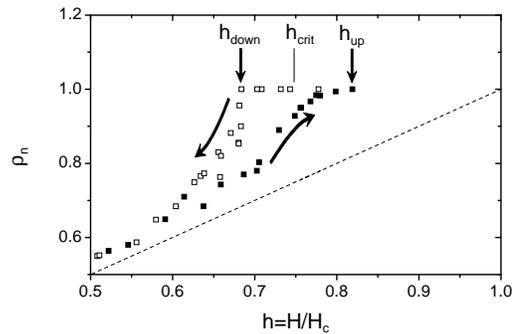}}
	\end{center}
	\caption{Area fraction of the normal phase $\rho_n$ as a function of the reduced applied magnetic field $h$ for the 10 $\mu$m thick sample at T=1.88 $\pm$ 0.04 K ($Bm$=3.2). Black squares: increasing field, empty squares: decreasing field, dashed line: $\rho_n$=$H/H_c$. The experimental nucleation ($h_{down}$) and collapse ($h_{up}$) fields are indicated by arrows below and above the calculated critical field $h_{crit}$, respectively.}
	\label{fig:rhon}
\end{figure}
The associated magnetic hysteresis is displayed in Fig. \ref{fig:rhon}. The area fraction of the NS phase $\rho_n$, determined from magneto-optical images, is plotted as a function of the reduced applied field $h=H/H_c$. The magnetic field $H_n$ in the NS domains is related to $\rho_n$ by flux conservation $\rho_n H_n=H$. There is a clear deviation from $\rho_n=H/H_c$ shown by the dashed line, which means that $H_n < H_c$. The transition to the NS ($\rho_n$=1) is thereby completed at a lower field than $H_c$.  However, the transition is not characterized by a unique transition field but by two fields $h_{up}$ and $h_{down}$. $h_{up}$ ($h_{down}$) are the fields at which the SC domains collapse (appear) when $h$ is increased (decreased).

\begin{figure}[htbp]
	\begin{center}
		\scalebox{0.45}{\includegraphics{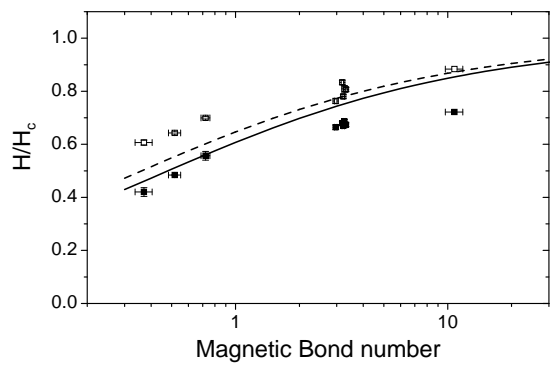}}
	\end{center}

	\caption{Experimental nucleation field $h_{down}$ (black squares) and collapse field $h_{up}$(empty squares) as a function of the magnetic Bond number. The full and dashed curves are the theoretical critical and collapse fields, respectively.}
	\label{fig:chp-nucleation-collapse}
\end{figure}

\begin{figure}[htbp]
	\begin{center}
		\scalebox{0.45}{\includegraphics{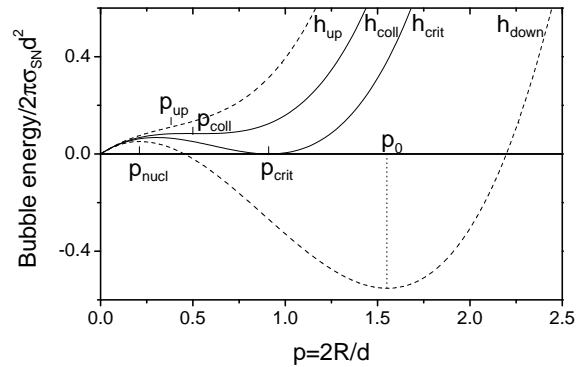}}
	\end{center}
	\caption{Energy of a single SC bubble in the NS matrix in units of $2\pi\sigma_{SN}d^2$ as a function of the reduced diameter. The $Bm$-value of the 10 $\mu$m thick film, Bm=3.2, is used.}
	\label{fig:energy}
\end{figure}

This magnetic hysteresis was observed in all the studied samples. In Fig. \ref{fig:chp-nucleation-collapse}, $h_{up}$ and $h_{down}$ are reported as a function of the magnetic Bond number $Bm=d/2\pi\Delta(T)$ \cite{jeudy}. $d$ is the sample thickness and $\Delta(T)$ is the interface wall parameter \cite{note+sharvin}. In order to determine whether $h_{up}$ and $h_{down}$ are related to superheating and supercooling, they were compared to the transition field deduced from the free energy of the system. In the framework of the CCL model, we calculated the free energy associated with the formation of an isolated SC cylindrical domain 
\begin{equation}
	F=2\pi \sigma_{SN} d^2\left[\frac{p}{2}+ Bm\left(h^2-1\right)\pi \frac{p^2}{4}+ Bm\: h^2\frac{p^3}{3}\right]\: ,
\label{CCLfree-energy}
\end{equation}
where $\sigma_{SN}=\Delta H_c^2/8\pi$ is the interfacial tension between the SC and NS phases, $p=2R/d$ is the reduced bubble diameter.
The first term in Eq. \ref{CCLfree-energy} is the interface energy. The second term contains the bulk magnetic energy and the condensation energy. It is negative since $h<1$. The third term represents the interaction energy of the screening current circulating within the bubble wall.
The bubble energy is plotted in Fig. \ref{fig:energy} as a function of $p$. The set of curves obtained for different applied fields presents the typical behavior of a metastable system. The critical field $h_{crit}=-x^{1/2}+\left(1+x\right)^{1/2}$, $x=8/\left(3\pi^2 Bm\right)$, is the field at which the free energies with and without a SC bubble are equal. An energy barrier impedes the nucleation or the collapse of a bubble at $h=h_{crit}$. Starting from the NS phase ($p=0$) and decreasing $H$, the system may stay in a metastable state. Nucleation of a SC bubble occurs if $p>p_{nucl}$ with $p_{nucl}=y\left(1-\left(1-8/z\right)^{1/2}\right)$, $y= \pi\left(1-h^2\right)/\left(4 h^2\right)$, $z=\pi^2\left(1-h^2\right)^2 Bm/h^2$; $h$ stands here for the nucleation field. The expansion of the SC phase is however limited by the long-range Biot-Savart interaction of the screening current ($p^3$ term in Eq. \ref{CCLfree-energy}) leading to an equilibrium bubble diameter $p_0=y\left(1+\left(1-8/z\right)^{1/2}\right)$.
In the following we assume that the first step of nucleation yields circular domains. Their evolution towards the laminar shape will be discussed later.

Starting from the IS state with SC bubbles and rising the field, the system may remain in a metastable state above $h_{crit}$, up to the collapse field corresponding to the disappearance of the energy barrier: $h_{coll}=-w^{1/2}+\left(1+w\right)^{1/2}$ with $w=2/(\pi^2 Bm)$. The corresponding collapse diameter is $p_{coll}=\left(1+\left(1+1/w\right)^{1/2}\right)/(\pi Bm)$.

\begin{figure}[htbp]
	\begin{center}
		\scalebox{0.40}{\includegraphics{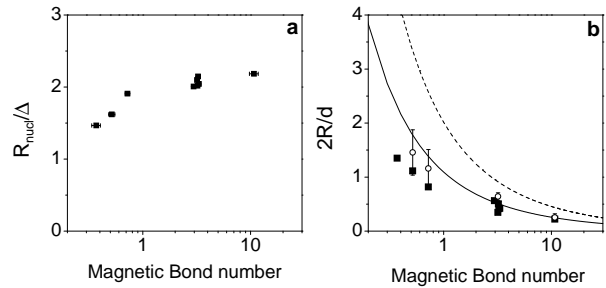}}
	\end{center}
	\caption{(a) Nucleation radius (in units of $\Delta$) obtained from the nucleation barrier at $h_{down}$. (b) calculated critical (dashed curve) and collapse (full curve) diameters, collapse diameter obtained from $h_{up}$ (black squares), measured SC bubble diameter close to $h_{up}$ (empty circles).}
	\label{fig:Rnucl-Rcoll}
\end{figure}

$h_{crit}$ and $h_{coll}$ are compared to $h_{down}$ and $h_{up}$ in Fig. \ref{fig:chp-nucleation-collapse}. $h_{down}$ and $h_{up}$-values lie below and above $h_{crit}$, respectively. This is consistent with the existence of barriers for nucleation and collapse. SCL and SH states are indeed observed. Since the $h_{down}$ fields remain much larger than the spinodal limit $H_{c2}/H_c\approx$ 0.12 for indium, the onset of the SC phase proceeds through heterogeneous nucleation. Thermally activated nucleation can be ruled out since the barrier is larger than the thermal energy $kT$ by many orders of magnitude. Defects likely act as potential wells that locally cancel the nucleation barrier when its amplitude is sufficiently lowered by the applied field. The nucleation radius $R_{nucl}=p_{nucl}d/2$ is obtained from the nucleation field $h_{down}$ and plotted in Fig. \ref{fig:Rnucl-Rcoll}(a) in units of $\Delta$. $R_{nucl}$ is of the order of 2 $\Delta$ showing a variation by only a factor 1.5 while $Bm$ is changed by a factor 30. The nucleation volume is thereby $\approx \pi \Delta^2 d$. This is quite reasonable since $\Delta$ is of the order of the coherence length $\xi$, which is the typical dimension of perturbation of the order parameter when a SC domain nucleates. A more accurate description of nucleation should take into account the spatial variation of the order parameter at the SC-NS interface, but this is beyond the scope of the CCL model.

Considering the collapse of SC bubbles, let us note that the $h_{up}$-values lie even above $h_{coll}$. As the movement of SC bubbles is frozen close to $h_{up}$, this shift of the SC-NS transition beyond the collapse field likely originates from the existence of pinning centers that form local potential wells. For all samples studied they are found to decrease the energy of the system by almost the same quantity as the potential wells that cancel the nucleation barrier. This suggests a common nature for nucleation and pinning centers.

Let us examine now whether the CCL model, which describes well the magnetic hysteresis, also provides a good agreement for domain sizes. In Fig. \ref{fig:Rnucl-Rcoll}(b) the average diameter of SC bubbles measured close to $h_{up}$ is compared to the calculated diameters $p_{crit}$ and $p_{coll}$ and to the diameter $p_{up}$.  $p_{up}$ corresponds to the p-value for which $\partial^2 F/\partial^2 p$=0 at $h=h_{up}$. The measured diameters are in quite good agreement with $p_{coll}$ and $p_{up}$, thereby indicating that the CCL model accurately describes domain sizes.
\begin{figure}[htbp]
	\begin{center}
		\scalebox{0.45}{\includegraphics{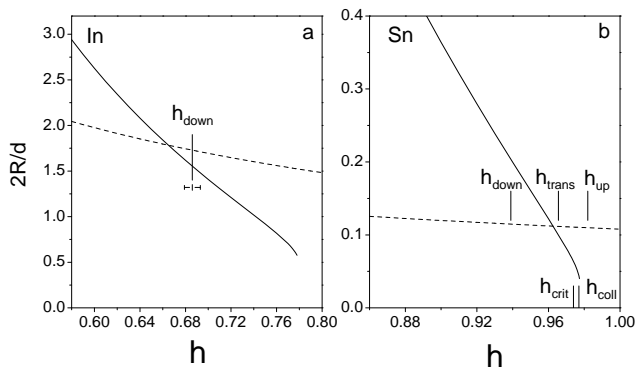}}
	\end{center}
	\caption{Equilibrium diameter of SC nucleated bubble (solid line) and critical diameter for elliptical instability (dashed line). (a) indium ($Bm$=3.2);(b) tin ($Bm$=387), experimental data from \cite{egorov}, $H_c$=305.5 G \cite{egorov}, $\Delta$=0.23 $\mu$m \cite{huebener}. }
	\label{fig:rayon-critique-instabilite}
\end{figure}

Let us now address the question of domain-shape hysteresis: why does nucleation of the SC phase give rise to the lamellar pattern even though the ground state of the system close to the critical field is the bubble phase? It was suggested that the ramification of the SC phase propagating in the NS phase originates from dynamical instabilities driven by magnetic field diffusion \cite{frahm,liu}. However the role of long-range interactions in branching instabilities was later emphasized \cite{langer,dorsey}. We consider here only the onset of domain formation. We show that shape instability arising from long-range magnetic interactions very likely accounts for topological hysteresis. From linear stability analysis \cite{cebers-janmey} the critical diameter for the bubble elliptical instability is obtained as $p_{inst}=3/h\sqrt{2 Bm}$. If the bubble diameter after nucleation $p_0$ is larger than $p_{inst}$ a bubble evolves into an elongated-shape domain. $p_0$ and $p_{inst}$ are plotted as a function of the field in Fig. \ref{fig:rayon-critique-instabilite}(a) for the 10 $\mu$m indium sample ($Bm$=3.2). At the nucleation field shown by the vertical bar the nucleated bubble diameter is very close to the instability limit, being smaller by only 10 \%. 

Moreover our theoretical predictions also provide excellent agreement with experimental data obtained by muon spin rotation spectroscopy on white tin \cite{egorov} as shown in Fig. \ref{fig:rayon-critique-instabilite}(b). The magnetic Bond number $Bm=387$ is much larger than for indium. The field of disappearance of the bubble phase ($h_{up}$) is slightly larger than the calculated collapse field $h_{coll}$. The crossing between $p_0$ and $p_{inst}$ coincides with the field $h_{trans}$ at which a transition from the lamellar to the bubble phase is observed in rising field. In decreasing field, nucleation occurs at $h_{down}$, below the crossing point. Nucleated bubbles are unstable and therefore the lamellar phase is observed. 

We propose the following description of the hysteretic SC-NS transition. In rising field, the SC lamellar phase evolves towards the bubble phase which can remain in a metastable state above the critical field. The complete transition to the normal phase occurs with the collapse of finite-size bubbles at the collapse field or slightly above if SC domains are trapped in local potential wells. In decreasing field, the NS phase is a metastable state. Nucleation occurs below the critical field due to the existence of a nucleation barrier. Depending on the magnetic Bond number and the value of the nucleation field, nucleation may yield unstable bubbles with respect to elliptical deformation. They evolve into lamellae with subsequent growth of the lamellar pattern in order to reach the equilibrium state corresponding to the applied field. 

In conclusion, the SC-NS phase transition in type-I SC films exhibits magnetic hysteresis and domain-shape hysteresis, which are shown to arise from different physical phenomena. Magnetic hysteresis, characterized by different values of the collapse and nucleation fields of SC domains, is found to be the signature of metastable states. Domain-shape hysteresis manifests itself as the collapse of bubble domains and nucleation of lamellar domains. Bubble-shape elliptical instability provides a very likely explanation for this topological hysteresis for a broad range of values of the magnetic Bond number. An analysis along the same lines would be of valuable interest for other systems with long-range interactions that exhibit similar domain patterns and hysteretic behavior.



\begin{thebibliography}{breitestes Label}
\bibitem{HKR}R.P. Huebener, R.T. Kampwirth, and V.A. Rowe, Cryogenics {\bf 12}, 100 (1972).
\bibitem{FHK}D.E. Farrell, R.P. Huebener, and R.T. Kampwirth, J. Low Temp. Phys. {\bf 19}, 99 (1975).
\bibitem{faber58}T.E.Faber, Proc.R.Soc. London Ser.A {\bf{248}}, 460 (1958).
\bibitem{cebersPRB}A. C\={e}bers, C. Gourdon, V. Jeudy, T. Okada, Phys. Rev. B {\bf 72}, 014513 (2005).
\bibitem{hubertschafer}A. Hubert and R. Sch\"{a}fer, \textit{Magnetic domains}, Springer, Berlin, 2000.
\bibitem{molho}P. Molho, J.L. Porteseil, Y. Souche, J. Gouzerh and J.C.S. Levy, J. Appl. Phys. {\bf 61}, 4188 (1987).
\bibitem{seul-chen}M. Seul and V.S. Chen, Phys. Rev. Lett. {\bf 70}, 1658 (1993).
\bibitem{bacri}J.C. Bacri, D. Salin, J. Phys. (Lettres) {\bf 43}, L771 (1982).
\bibitem{langer}S.A. Langer, R.E. Goldstein, and D.P. Jackson, Phys. Rev. A {\bf{46}}, 4894 (1992).
\bibitem{dorsey}A.T. Dorsey and R.E. Goldstein, Phys. Rev. B {\bf 57}, 3058 (1998).
\bibitem{cebers2}A. Cebers, Magnetohydrodynamics {\bf 37}, 195 (2001).
\bibitem{frahm}H. Frahm, S. Ullah, and A.T. Dorsey, Phys. Rev. Lett. {\bf 23}, 3067 (1991).
\bibitem{liu}Fong Liu,M. Mondello, and N. Goldenfeld, Phys. Rev. Lett. {\bf 66}, 3071 (1991).
\bibitem{onuki}H. Kitamura, and A. Onuki, arXiv:cond-mat 0508030 v1 1 Aug 2005.
\bibitem{tinkham}M. Tinkham, \textit{Introduction to Superconductivity} p. 95, McGraw-Hill, New York, 1975.
\bibitem{jeudy}V. Jeudy, C. Gourdon, T. Okada, Phys. Rev. Lett. {\bf 92}, 147001 (2004).
\bibitem{egorov}V.S. Egorov, G. Solt, D. Herlach, and U. Zimmermann, Phys. Rev. B \textbf{64}, 024524 (2001).
\bibitem{feder}J. Feder and D.S. McLachlan, Phys. Rev. \textbf{177}, 763 (1969).
\bibitem{smith}F.W. Smith, A. Baratoff and M. Cardona, Phys. Kond. Materie \textbf{12}, 145 (1970).
\bibitem{gourdon}C. Gourdon \textit{et al.}, Appl. Phys. Lett. {\bf 82}, 230 (2003).
\bibitem{note+sharvin} We use $\Delta(T) =\Delta(0)/\sqrt{1-T/T_c}$ according to Yu. V. Sharvin, Soviet Phys. JETP {\bf 11}, 216 (1960), with $T_c$=3.4 K and $\Delta (0) =
0.33\: \mu$m for indium.
\bibitem{huebener}R.P. Huebener, \textit{Magnetic Flux structures  in
Superconductors}, 2nd ed., p. 10, Springer Verlag, New York, 2001.
\bibitem{cebers-janmey}A. Cebers and P.A. Janmey, J. Phys. Chem. B \textbf{106}, 12351 (2002).


\end{thebibliography}
\end{document}